\documentclass[aps,prd,twocolumn,showpacs,superscriptaddress]{revtex4-1}

\usepackage{graphicx}
\usepackage{amsmath,amssymb,bm}
\usepackage{multirow}
\usepackage{float}
\usepackage{hyperref}
\usepackage{booktabs}
\hypersetup{colorlinks=true,linkcolor=blue,citecolor=blue,urlcolor=blue}
\newcommand{\orcid}[1]{%
  \href{https://orcid.org/#1}{%
    \includegraphics[height=2ex, keepaspectratio]{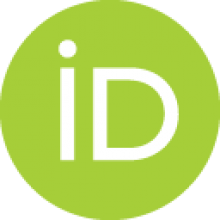}%
  }%
}

\begin{document}

\title{Identifying lensed gravitational waves with physics-informed posterior learning}

\author{Tian-Yang Sun\orcid{0009-0002-5109-6420}}
\affiliation{Liaoning Key Laboratory of Cosmology and Astrophysics, College of Sciences, Northeastern University, Shenyang 110819, China}

\author{Xiao Guo\orcid{0000-0001-5174-0760}}
\affiliation{School of Physics Science and Technology, Wuhan University, Wuhan 430072, China}

\author{Jing-Fei Zhang\orcid{0000-0002-3512-2804}}
\affiliation{Liaoning Key Laboratory of Cosmology and Astrophysics, College of Sciences, Northeastern University, Shenyang 110819, China}

\author{Xin Zhang\orcid{0000-0002-6029-1933}}
\thanks{Corresponding author}
\email{zhangxin@neu.edu.cn}
\affiliation{Liaoning Key Laboratory of Cosmology and Astrophysics, College of Sciences, Northeastern University, Shenyang 110819, China}
\affiliation{MOE Key Laboratory of Data Analytics and Optimization for Smart Industry, Northeastern University, Shenyang 110819, China}
\affiliation{National Frontiers Science Center for Industrial Intelligence and Systems Optimization, Northeastern University, Shenyang 110819, China}

\begin{abstract}
Gravitational lensing of gravitational waves can probe compact lenses, dark matter substructure, and cosmological distances, but identifying lensed events is difficult when unrelated binary mergers overlap in the same analysis window. We develop physics-informed posterior learning for ranking lensed multi-image signals against unrelated multiple-merger events. The method exploits the geometric-optics consistency that lensing can change amplitudes, arrival times, and Morse phase offsets while preserving the intrinsic phase evolution of the source. We infer a simulation-trained approximate posterior for the common detector-frame chirp mass and symmetric mass ratio, and fuse posterior samples with direct waveform features. Training uses generic multi-image simulations, while point-mass, singular-isothermal-sphere, singular-isothermal-ellipsoid, and shear-perturbed lenses are reserved for held-out lens-family evaluation. For the observationally motivated binary-black-hole population, the fusion ranking raises the detection efficiency from $20.8\%$ to $35.2\%$ at a $1\%$ reference false-positive-rate threshold calibrated on the corresponding unrelated multiple-merger sample. It lowers the network signal-to-noise ratio needed for $50\%$ detection efficiency from 45.3 to 33.5, which corresponds to a 1.35 times larger signal-to-noise-ratio-equivalent distance scale. The gain is limited by loud unrelated multiple-merger events that are partly source consistent, and by the need to calibrate the unrelated multiple-merger population. These results suggest that physical consistency can become a guiding principle for machine learning searches in dense gravitational-wave catalogs.
\end{abstract}

\maketitle

\section{Introduction}

The detection of gravitational waves (GWs) from compact-binary coalescences has opened a way to test gravity, measure compact-object populations, and use standard sirens for cosmology \cite{Schutz:1986gp,LIGOScientific:2016lio,LIGOScientific:2017adf,Chen:2017rfc,LIGOScientific:2018dkp,Wang:2018lun,Zhang:2018byx,LIGOScientific:2019fpa,Zhang:2019ple,Zhang:2019loq,Wang:2019tto,Zhao:2019gyk,Jin:2020hmc,LIGOScientific:2020tif,Biscoveanu:2020are,Roulet:2021hcu,Jin:2021pcv,Gong:2021jgg,Biscoveanu:2022qac,Fishbach:2022lzq,Callister:2022qwb,Tong:2022iws,KAGRA:2021duu,LIGOScientific:2021aug,Jin:2022qnj,Song:2022siz,Jin:2023sfc,Sun:2023bvy,LIGOScientific:2021sio,Song:2025ddm,Jin:2025dvf,Song:2025bio,Du:2025odq,Song:2026kii,Xiong:2026piv,Song:2026tdy,Du:2026jtf}.
As detector sensitivity improves, gravitational lensing of GWs becomes a direct target \cite{ET:2019dnz,Barausse:2020rsu,Ruan:2018tsw,Evans:2021gyd,Luo:2025ewp}.
Lensed events can provide repeated views of the same merger, constrain compact lenses and small dark matter structures, and improve the use of dark binary-black-hole events as cosmological probes \cite{Sereno:2010dr,Dai:2017huk,Li:2018prc,Ng:2017yiu,Lai:2018rto,Liu:2019dds,Cao:2019kgn,Gais:2022xir,Guo:2022dre,Xu:2021bfn,Wang:2021kxc,Liu:2021xvc,Cao:2022mrc,Huang:2023prq,Tambalo:2022wlm,Meena:2023qdq,Cheung:2024ugg,Smith:2025axx,Chen:2025xeg,Wu:2026ccw,Ando:2026poq,Ando:2026eam}.
The first step toward this science is not the final measurement of a lens system.
It is the construction of an event ranking that remains useful when the lens potential, the number of images, the magnifications, and the time delays are not known in advance.
This step is especially important when multiple compact-binary signals fall in the same analysis window, where a lensed multi-image signal can be confused with unrelated mergers in the same window.

No unambiguous lensed GW event has yet been established.
Current searches approach the problem from several complementary directions.
Some methods compare pairs or groups of catalog events through posterior overlap, time-delay consistency, sky localization, magnification, and merger-rate information \cite{Haris:2018vmn,Janquart:2021qov,LIGOScientific:2021izm,Lo:2021nae,Janquart:2023mvf,Caliskan:2022wbh,Liu:2023ikc,LIGOScientific:2023bwz,Goyal:2023lqf,Barsode:2024zwv,Shan:2023ngi,Su:2025xry,LIGOScientific:2025cwb,Chan:2025kyu,Hannuksela:2025wgv,Barsode:2025agk,Shan:2025jpt,Kopty:2026ugr,Heynen:2025fgq,Chakraborty:2025lmk,Chakraborty:2026yci,Mould:2026hfh,Chakraborty:2025pxt}.
Other methods search for single-event signatures, including microlensing, diffraction, parity-related waveform changes, and time-frequency distortions \cite{Fernandez-Nunez:2016urh,Fernandez-Nunez:2016cea,Dai:2018enj,Diego:2019lcd,Jung:2017flg,Diego:2019rzc,Guo:2020eqw,Wang:2021kzt,Cheung:2020okf,Mishra:2021xzz,Choi:2021bkx,Urrutia:2021qak,Wang:2021lij,Seo:2021psp,Goyal:2023uvm,Tambalo:2022plm,Caliskan:2023zqm,Fairbairn:2022xln,Brando:2024inp,Jana:2024dhc}.
These studies define the physical checks that a convincing lensed event should pass.
They also show why the ranking problem can remain open before detailed follow-up.
A signal may be visible in the strain data while its lens model is uncertain, and an unrelated multiple-merger event can bias parameter estimation or mimic part of the repeated-image morphology \cite{Relton:2021cax,Keitel:2024brp,Diao:2025hrl,Rao:2025poe}.
The recent discussion of the fourth observing run (O4) outlier GW231123\_135430 makes this ambiguity concrete.
The same data have motivated interpretations in terms of overlapping signals and lensing-related waveform distortions \cite{Goyal:2025eqo,Hu:2025lhv,Wang:2026yjk}.
This is the regime addressed here.
The goal is not to perform final lens modeling for a confirmed event, but to rank lensed multi-image signals against physically relevant unrelated multiple-merger events.

Deep learning has become useful for classification and fast inference in GW data analysis and cosmology \cite{DBLP:conf/cvpr/HeZRS16,Zevin:2016qwy,Gabbard:2017lja,George:2016hay,George:2017pmj,Xia:2020vem,Chatterjee:2021lit,Whittaker:2022pkd,Schafer:2022dxv,Zhao:2022qob,Sun:2023vlq,Wang:2023lif,Wang:2022quo,Xu:2024jbo,Shi:2024age,Wang:2024oei,Sun:2026wep}.
Machine-learning classifiers have also been applied to GW lensing searches, where they can quickly rank events from waveform morphology or learned image-like representations \cite{Goyal:2021hxv,Kim:2022lex,Magare:2024wje,Offermans:2024fot,Wang:2025loo,Liu:2025ixi,Campailla:2025ftj,Zhang:2026gdt,Li:2025ngt}.
Such classifiers are attractive because they are fast after training and can use information that is difficult to compress into a small set of hand-built statistics.
Their score, however, need not be controlled by the source-consistency physics that makes a set of lensed images different from an unrelated multiple-merger event.
At the other end, full Bayesian lensing inference is more interpretable but can be too expensive when the lens model and image number are not fixed.
Simulation-based inference offers a middle route.
Neural posterior estimators can provide fast approximate posteriors for GW parameter estimation, lensing inference, population studies, and cosmological applications \cite{DBLP:conf/icml/RezendeM15,DBLP:journals/corr/PapamakariosPM17,DBLP:conf/iclr/DinhSPL19,DBLP:conf/icml/ZieglerR19,DBLP:conf/nips/DurkanB0P19,Green:2020hst,Green:2020dnx,Dax:2021tsq,Williams:2021qyt,Gabbard:2019rde,Shen:2019vep,Dax:2022pxd,Leyde:2023iof,Du:2023plr,Sun:2023vlq,Dax:2024mcn,Bada-Nerin:2024wkn,DBLP:journals/corr/abs-2505-10466,Xiong:2024gpx,Sun:2024ywb,Sun:2025ypd,Qin:2025mvj,Sun:2026dga}.
Their uncertainty can be checked with posterior diagnostics under the assumed simulation model \cite{DBLP:journals/corr/abs-2110-06581}.

We use lensing source consistency as a physical intermediate representation for classification.
In geometric optics, lensing changes image amplitudes, arrival times, and Morse phase offsets, while preserving the intrinsic frequency evolution of the source \cite{1986ApJ...307...30D,1992grle.book.....S,Takahashi:2003ix,Takahashi:2004mc,2006glsw.conf.....M,Meena:2019ate,Bulashenko:2021fes,Bondarescu:2022srx,Ali:2022guz}.
For compact binaries, detector-frame chirp mass and symmetric mass ratio form a low-dimensional phase-evolution diagnostic.
Lensed images should be compatible with one common source in this mass plane, whereas unrelated multiple-merger events need not be.
We train a neural posterior estimator (NPE) for these common-source parameters and pass finite posterior samples to a permutation-invariant classifier.
The classifier is trained with generic multi-image simulations.
Point mass (PM), singular isothermal sphere (SIS), singular isothermal ellipsoid (SIE), and SIE plus external shear (SIE+Shear) lenses are used only for held-out lens-family evaluation.
This separation between training and physical lens-family testing is the sense in which the ranking task tests lens-model generalization.
It tests whether posterior samples can carry source-consistency information beyond the particular lens families used for evaluation.

The organization of this paper is as follows.
Section~\ref{sec:physical_setup} describes the data simulation, learning task, network architecture, classification metrics, and posterior diagnostics.
Section~\ref{sec:results} presents the results and discussion.
Section~\ref{sec:conclusion} summarizes the conclusions.

\section{Method}
\label{sec:physical_setup}

\subsection{Data simulation principle}

In the geometric-optics limit, a lensed frequency-domain waveform can be written as a coherent sum over images,
\begin{equation}
    \tilde{h}_{\rm L}(f)
    =
    \sum_{j=1}^{N_{\rm img}}
    |\mu_j|^{1/2}
    \exp\left[
        2\pi i f \Delta t_j
        - i \pi n_j
    \right]
    \tilde{h}_{\rm src}(f;\bm{\theta}_{\rm s}),
    \label{eq:lensed_waveform}
\end{equation}
where \(\tilde{h}_{\rm src}\) is the unlensed source waveform, \(\mu_j\) is the signed magnification, \(\Delta t_j\) is the time delay, and \(n_j\) labels the Morse phase.
In the convention used here, \(n_j=0,1/2,1\) for type-I, type-II, and type-III images, so the image phase factor is \(\exp(-i\pi n_j)\).
Equivalently, one may use an integer Morse index \(N_j=0,1,2\) and write the same factor as \(\exp(-i\pi N_j/2)\).
The image-level nuisance parameters \((\Delta t_j,\mu_j,n_j)\) depend on the lens potential and describe how images are arranged.
The discriminating information is whether the arrivals remain consistent with one shared \(\bm{\theta}_{\rm s}\) after these lensing transformations are allowed.

The full source vector \(\bm{\theta}_{\rm s}\) contains masses, spins, sky position, inclination, distance, coalescence time, phase, and other waveform parameters.
We do not infer this full vector.
Instead, we use a two-dimensional phase-consistency projection,
\begin{equation}
    \bm{\vartheta}_{\rm ph} =
    g(\bm{\theta}_{\rm s}) =
    \left(\log \mathcal{M}_{c,z},\eta\right),
    \qquad
    \eta = \frac{m_{1,z}m_{2,z}}{(m_{1,z}+m_{2,z})^2}.
    \label{eq:phase_projection}
\end{equation}
The detector-frame chirp mass is
\begin{equation}
    \mathcal{M}_{c,z} =
    \frac{(m_{1,z}m_{2,z})^{3/5}}{(m_{1,z}+m_{2,z})^{1/5}}.
\end{equation}
Here \(m_{i,z}\) denotes the detector-frame mass, related to the source-frame mass by \(m_{i,z}=(1+z)m_{i,{\rm src}}\).
The symmetric mass ratio satisfies \(0<\eta\le 1/4\).
The upper limit corresponds to equal masses, while smaller values describe more unequal binaries.
This lower-dimensional projection retains the mass-plane phase-consistency information needed for the O4 LIGO Hanford and Livingston (H1/L1) analysis while reducing the data demand of neural posterior estimation.
Other source parameters, including sky position, inclination, distance, and coalescence phase, are generated and stored for diagnostics.
Sky position does not add intrinsic frequency-evolution information.
It enters through detector time delays and antenna-pattern projection, and is therefore treated separately as a detector-response consistency extension rather than as part of the main phase-consistency target.
In this controlled nonspinning binary black hole (BBH) study, spin effects are left to higher-dimensional source-posterior extensions.

\subsection{Dataset construction and learning problem}
\label{sec:data}

The input to all classifiers is a two-detector whitened time series \(\mathbf{x} \in \mathbb{R}^{2\times 8192}\), corresponding to H1 and L1 at a sampling frequency of \(4096\,{\rm Hz}\).
Signals are injected into pre-vetoed O4 H1/L1 Gravitational Wave Open Science Center (GWOSC) background segments stored in the data-generation file \( {\tt O4\_background\_vetoed.h5} \) \cite{LIGOScientific:2025snk}.
The file contains 94 strain segments from GPS \(1382928002\) to \(1385284831\), with a total duration of 15 days.
The stored channels are H1:GWOSC-16KHZ\_R1\_STRAIN and L1:GWOSC-16KHZ\_R1\_STRAIN.
We therefore identify the background as the vetoed segments used by the simulation pipeline, rather than assigning an additional data-quality flag that is not stored in the HDF5 metadata.
For each example a \(4\,{\rm s}\) segment is used for injection and whitening, and the central \(2\,{\rm s}\) window is retained.
The waveform model used for injections is IMRPhenomXPHM with a low-frequency cutoff of \(20\,{\rm Hz}\), within the broader compact-binary waveform and parameter-estimation ecosystem used in ground-based GW analyses \cite{Ajith:2009bn,Veitch:2014wba,Harry:2016ijz,Biwer:2018osg,Estelles:2021jnz,Mateu-Lucena:2021siq,Estelles:2021gvs}.
Although this waveform approximant supports precessing-spin effects, the injected population in this controlled study is nonspinning.
The main physical and injection parameters are summarized in Table~\ref{tab:simulation_parameters}.

The simulated sources are BBH coalescences.
The lensed training class consists of generic lensed multi-image signals.
For each example the number of images is sampled uniformly from 2 to 5.
The image amplitude factors are sampled in the range 1--20 and sorted from large to small.
The relative arrival times are sampled within \([-0.3,0.3]\,{\rm s}\), and the Morse phase index is sampled from \(\{0,1/2,1\}\).
These simulations are not labeled by a physical lens family during classifier training.
The generic image prior only sets an envelope over multiplicity, amplitude, time delay, and Morse phase, not a physical lens-population rate model.
The PM, SIS, SIE, and SIE+Shear tests below provide the corresponding held-out lens-family check.
The unrelated multiple-merger training class is formed by superposing 2--5 unrelated unlensed compact-binary signals in the same analysis window.
Each unrelated multiple-merger event has independent masses, distance, sky location, orientation, and phase, with coalescence times drawn from the same analysis window as the lensed images.
The task is therefore a controlled test of common-source consistency rather than multiplicity alone.

The labels are \(y=1\) for lensed signals and \(y=0\) for unrelated multiple-merger events.
Training uses cyclic data generation and loading, so successive training cycles expose the model to different noise realizations and physical configurations rather than one fixed dataset.
The signal-to-noise-ratio (SNR)-matched run draws from independently seeded lensed and unrelated multiple-merger files throughout training.
The network SNR is
\begin{equation}
    \rho_{\rm net}
    =
    \left(\sum_I \rho_I^2\right)^{1/2},
    \qquad
    \rho_I^2
    =
    4\int_{f_{\rm low}}^\infty
    \frac{|\tilde{h}_I(f)|^2}{S_{n,I}(f)}\,{\rm d} f ,
\end{equation}
where \(I\) labels detectors and \(S_{n,I}(f)\) is the one-sided noise power spectral density.

The test sets are generated separately from PM, SIS, SIE, and SIE+Shear lenses and are not used as classifier labels during training.
For each lens family and SNR, all three classifiers are evaluated on the same independently generated lensed events and unrelated multiple-merger events, giving 28 held-out test conditions.
Separate diagnostic samples probe posterior morphology, unrelated multiple-merger tails, score changes across lens family and SNR, and population-shift behavior under an observationally motivated BBH mass distribution.

\begin{table*}[t]
\caption{
Physical and injection parameters used to construct the multi-signal data.
The NPE target is the phase-consistency projection \(\bm{\vartheta}_{\rm ph}=g(\bm{\theta}_{\rm s})\), not the full source vector \(\bm{\theta}_{\rm s}\).
}
\label{tab:simulation_parameters}
\begin{ruledtabular}
\begin{tabular}{p{0.23\textwidth}p{0.15\textwidth}p{0.31\textwidth}p{0.23\textwidth}}
Parameter & Symbol & Prior/range & Sampling \\
\hline
Source-frame component masses & \(m_{1,{\rm src}},m_{2,{\rm src}}\) & \(5M_\odot\)--\(80M_\odot\), ordered as \(m_{1,{\rm src}}\ge m_{2,{\rm src}}\) & Independent uniform \\
Luminosity distance & \(D_L\) & \(10^2\)--\(10^{3.5}\,{\rm Mpc}\) & Log-uniform \\
Right ascension & \(\alpha\) & \([0,2\pi)\) & Uniform \\
Declination & \(\delta\) & \([-\pi/2,\pi/2]\) & Uniform in \(\sin\delta\) \\
Polarization angle & \(\psi\) & \([0,\pi)\) & Uniform \\
Inclination & \(\iota\) & \([0,\pi]\) & Uniform in \(\cos\iota\) \\
Coalescence phase & \(\phi_c\) & \([0,2\pi)\) & Uniform \\
Dimensionless spins & \(\chi_{1,2}\) & 0 & Fixed \\
Network SNR & \(\rho_{\rm net}\) & \(8\)--\(48\) & Rescaled after injection \\
\hline
Number of images/events & \(N\) & \(2\)--\(5\) & Discrete uniform \\
Image amplitude factor & \(|\mu_j|^{1/2}\) & \(1\)--\(20\) & Uniform\\
Relative time offset & \(\Delta t_j\) & \([-0.3,0.3]\,{\rm s}\) & Uniform\\
Morse phase index & \(n_j\) & \(\{0,1/2,1\}\) & Discrete \\
\end{tabular}
\end{ruledtabular}
\end{table*}

\subsection{Model architecture and training}
\label{sec:method}

\begin{figure*}[t]
\centering
\includegraphics[width=0.95\textwidth]{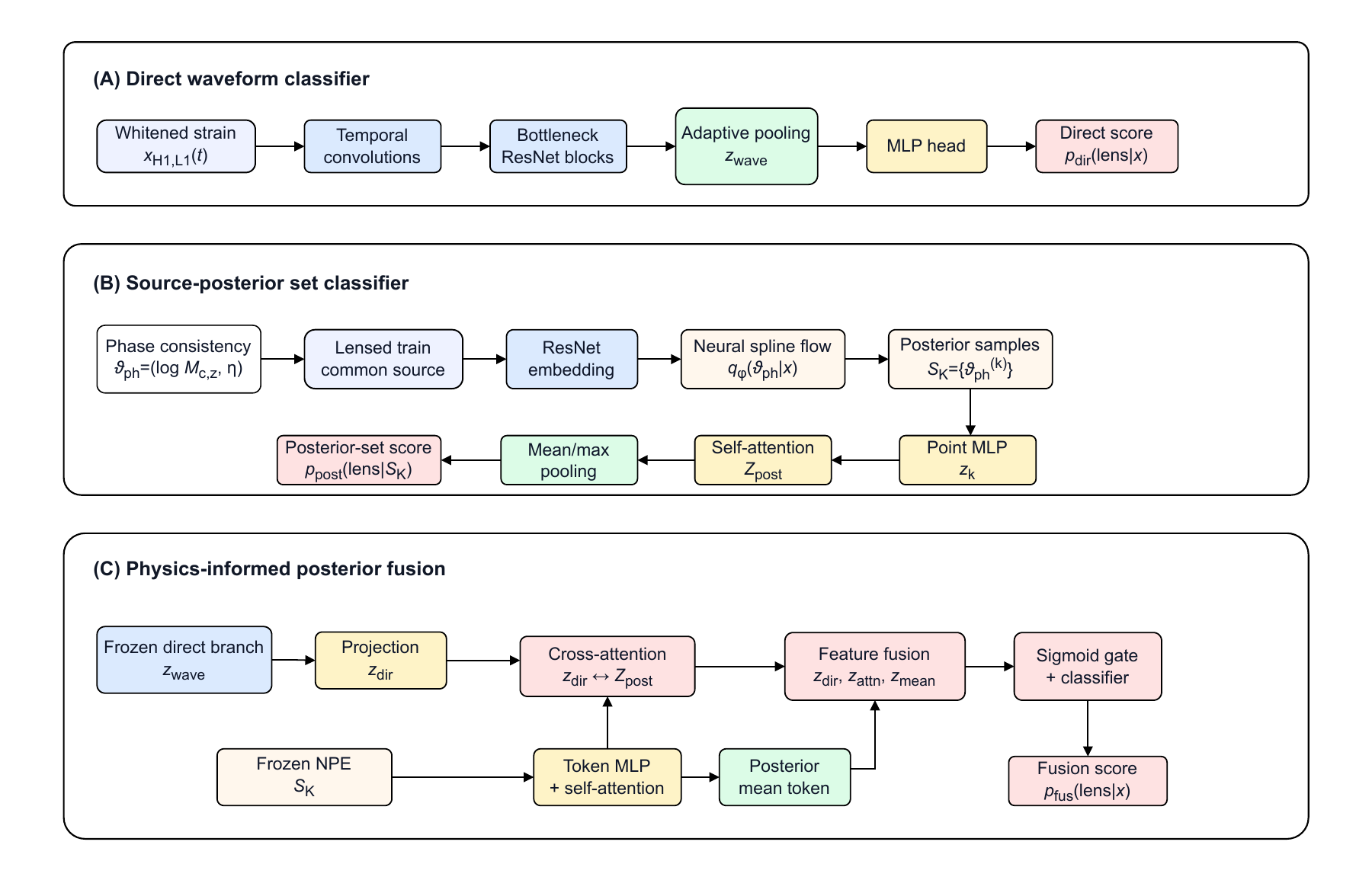}
\caption{
Architecture of the physics-informed posterior learning pipeline.
(A) The direct branch ranks events from waveform morphology alone.
(B) The posterior branch infers the simulation-trained approximate density \(q_\phi(\bm{\vartheta}_{\rm ph}|\mathbf{x})\) and treats finite posterior samples as an unordered set.
(C) The fusion branch keeps the frozen waveform and posterior representations separate until attention and gating combine them into a final ranking score.
The classifier is trained with generic multi-image simulations and tested on physical PM, SIS, SIE, and SIE+Shear lenses.
}
\label{fig:method_architecture}
\end{figure*}

Figure~\ref{fig:method_architecture} summarizes the three branches used in the analysis.

\subsubsection{Direct waveform classifier}

The direct classifier is a one-dimensional residual network that maps the whitened H1--L1 time series directly to a binary label.
It is used as the primary baseline because it measures the performance available from waveform morphology alone.
The backbone begins with a one-dimensional convolution with kernel size 7 and stride 2, followed by max pooling.
It then uses bottleneck residual blocks, each containing \(1\times1\), \(3\times1\), and \(1\times1\) convolutions with batch normalization and ReLU activations.
When the channel number or temporal resolution changes, the residual path uses a \(1\times1\) convolution for matching.
The channel sequence grows through 64, 256, 512, 1024, and 2048 channels, and adaptive average pooling produces a 2048-dimensional waveform feature.
The classification head is a fully connected \(2048\rightarrow512\rightarrow2\) network with ReLU and dropout.
No posterior information is supplied to this model, and its output score is interpreted as a ranking statistic for \(p(y=1|\mathbf{x})\).

\subsubsection{Neural posterior estimator}

The posterior estimator maps the same input \(\mathbf{x}\) to a simulation-trained approximate common-source conditional density over the two-dimensional phase-consistency parameter \(\bm{\vartheta}_{\rm ph}\),
\begin{equation}
    q_\phi(\bm{\vartheta}_{\rm ph}|\mathbf{x}) \simeq p(\bm{\vartheta}_{\rm ph}|\mathbf{x},\mathcal{H}_{\rm common}).
\end{equation}
Here \(\mathcal{H}_{\rm common}\) denotes the low-dimensional common-source hypothesis under the simulation prior.
It consists of a residual waveform encoder and a neural spline flow decoder.
The waveform encoder uses a one-dimensional ResNet-like architecture, reduces the temporal dimension to 16 positions with adaptive pooling, and then applies temporal attention pooling to produce the conditioning embedding.
The normalizing flow is a neural spline flow with a 16-dimensional conditioning embedding, a two-dimensional target, 9 invertible transforms, and hidden layers of width 4096.
The estimator is trained by minimizing the negative log likelihood
\begin{equation}
    \mathcal{L}_{\rm NPE}(\phi)
    =
    -\frac{1}{N}\sum_{i=1}^{N}
    \log q_\phi(\bm{\vartheta}_{{\rm ph},i}|\mathbf{x}_i).
    \label{eq:npe_loss}
\end{equation}
The optimizer is AdamW with learning rate \(2\times 10^{-4}\), weight decay \(10^{-4}\), cosine learning-rate annealing, and gradient clipping.
The best validation checkpoint is used for all posterior-set and fusion evaluations.

Only lensed examples train this posterior estimator, because unrelated multiple-merger events have no single true common-source parameter.
At classification time the estimator is applied to both classes, allowing the classifier to use the approximate-posterior structure produced when the common-source hypothesis is strained.

\subsubsection{Posterior-set classifier}

For each input waveform, the posterior estimator generates \(K\) samples,
\begin{equation}
    \mathcal{S}_K(\mathbf{x})=
    \{\bm{\vartheta}_{\rm ph}^{(1)},\ldots,\bm{\vartheta}_{\rm ph}^{(K)}\},
    \qquad
    \bm{\vartheta}_{\rm ph}^{(k)}\sim q_\phi(\bm{\vartheta}_{\rm ph}|\mathbf{x}).
\end{equation}
The posterior-set classifier treats \(\mathcal{S}_K\) as an unordered set.
Each posterior sample is first passed through a pointwise multilayer perceptron \(2\rightarrow128\rightarrow256\rightarrow512\), with batch normalization and ReLU activations.
The resulting sample features are aggregated with permutation-invariant max and mean pooling, concatenated into a 1024-dimensional vector, projected to a 512-dimensional feature vector, and classified by a \(512\rightarrow256\rightarrow2\) fully connected head with ReLU and dropout.
This is a Deep Sets-type architecture \cite{DBLP:conf/nips/ZaheerKRPSS17}.
It allows the model to use the posterior width and shape without depending on the ordering of posterior samples.

During the SNR-matched low-memory training run the classifier uses \(K=4\) posterior samples per example and a batch size of 8.
At evaluation, the posterior branches are evaluated with 32 posterior samples per example.
The classifier is trained with cross entropy using AdamW with learning rate \(10^{-4}\), weight decay \(10^{-4}\), cosine annealing, and gradient clipping.

\subsubsection{Attention-gated fusion classifier}

The fusion classifier combines the frozen 2048-dimensional waveform feature with posterior samples from the frozen estimator.
Posterior samples are embedded as tokens, refined by four-head self-attention, and combined with the direct waveform feature through four-head cross-attention.
The projected waveform feature, cross-attended posterior feature, and mean posterior-token summary are concatenated into a 768-dimensional vector, gated by a learned sigmoid gate, and passed to a \(768\rightarrow256\rightarrow2\) classifier.
The gate lets source-consistency information modify the ranking while retaining waveform evidence when posterior inference is noisy.

The SNR-matched fusion model is trained for 100 epochs with batch size 4 and \(K=4\) posterior samples per example.
The learning rate, weight decay, scheduler, and gradient clipping are the same as in the posterior-set classifier.
The final evaluation uses the best validation checkpoint and \(K=32\) posterior samples.

\subsection{Evaluation metrics and posterior diagnostics}
\label{sec:metrics}

Each classifier returns a lensing score \(s(\mathbf{x})\).
For a threshold \(\tau\), predictions with \(s\ge\tau\) are labeled as lensed.
The numbers of true positives, false positives, true negatives, and false negatives are denoted by TP, FP, TN, and FN.
The true-positive rate (TPR), recall, false-positive rate (FPR), and precision are
\begin{equation}
\begin{aligned}
    {\rm TPR}(\tau)
    &=
    {\rm Recall}(\tau)
    =
    \frac{{\rm TP}(\tau)}{{\rm TP}(\tau)+{\rm FN}(\tau)},\\
    {\rm FPR}(\tau)
    &=
    \frac{{\rm FP}(\tau)}{{\rm FP}(\tau)+{\rm TN}(\tau)},\\
    {\rm Precision}(\tau)
    &=
    \frac{{\rm TP}(\tau)}{{\rm TP}(\tau)+{\rm FP}(\tau)}.
\end{aligned}
\label{eq:classification_metrics}
\end{equation}
The receiver-operating-characteristic area under the curve (ROC-AUC) is the area under the \({\rm TPR}\) versus \({\rm FPR}\) curve as \(\tau\) is varied.
The precision--recall area under the curve (PR-AUC) is the area under the precision versus recall curve.
Recall at a fixed FPR, written as \(R@\alpha{\rm \ FPR}\), is \({\rm TPR}(\tau_\alpha)\) at a threshold \(\tau_\alpha\) chosen so that \({\rm FPR}(\tau_\alpha)=\alpha\).
ROC-AUC measures global ranking quality, while PR-AUC is reported only as a balanced-set diagnostic because realistic lensing searches have highly imbalanced class priors.
The fixed-FPR recalls provide the search-like operating points and are controlled by the high-score tail of the unrelated multiple-merger distribution.
For each model and lens family, table values are averaged over the sampled SNR range.
SNR-dependent figures are averaged over the four lens families at fixed SNR.

For the posterior estimator, we use two reliability diagnostics for the approximate common-source density \(q_\phi(\bm{\vartheta}_{\rm ph}|\mathbf{x})\) \cite{DBLP:journals/corr/abs-2110-06581}.
For a nominal credibility level \(\gamma\), the highest-posterior-density (HPD) credible region is
\begin{equation}
    \begin{aligned}
    C_\gamma(\mathbf{x})
    &=
    \{\bm{\vartheta}_{\rm ph}\,|\,q_\phi(\bm{\vartheta}_{\rm ph}|\mathbf{x})\ge \kappa_\gamma(\mathbf{x})\},
    \\
    \int_{C_\gamma(\mathbf{x})}
    q_\phi(\bm{\vartheta}_{\rm ph}|\mathbf{x})\,{\rm d}\bm{\vartheta}_{\rm ph}
    &=
    \gamma ,
    \end{aligned}
\end{equation}
where \(\kappa_\gamma\) is the density threshold that encloses probability \(\gamma\).
The empirical HPD coverage is
\begin{equation}
    \widehat{C}(\gamma)
    =
    \frac{1}{N}\sum_{i=1}^N
    \mathbf{1}\!\left[
    \bm{\vartheta}_{{\rm ph},i}^{\rm true}\in C_\gamma(\mathbf{x}_i)
    \right].
\end{equation}
A calibrated joint posterior should give \(\widehat{C}(\gamma)\simeq\gamma\).
For each one-dimensional marginal \(d\in\{\log \mathcal{M}_{c,z},\eta\}\), we also compute the probability-integral-transform value
\begin{equation}
    u_{i,d}
    =
    F_{i,d}\!\left(\vartheta_{i,d}^{\rm true}\right),
\end{equation}
where \(F_{i,d}\) is the marginal cumulative distribution function of \(q_\phi(\bm{\vartheta}_{\rm ph}|\mathbf{x}_i)\).
The probability--probability plot compares the empirical cumulative distribution of \(\{u_{i,d}\}\) with the diagonal uniform distribution, and the Kolmogorov--Smirnov (KS) statistic is
\begin{equation}
    D_d
    =
    \sup_{u\in[0,1]}
    \left|
    \widehat{F}_d(u)-u
    \right| .
\end{equation}

\section{Results and discussion}
\label{sec:results}

\subsection{Classification performance and held-out lens-family generalization}

Averaged over PM, SIS, SIE, and SIE+Shear and over the sampled SNR range, the direct waveform classifier reaches ROC-AUC \(0.785\) and PR-AUC \(0.753\).
The attention-gated fusion classifier reaches ROC-AUC \(0.844\) and PR-AUC \(0.805\), improving the direct baseline by 0.059 and 0.053, respectively.
At the stricter \(0.1\%\) false-positive-rate operating point, the recall increases from 0.020 for the direct classifier to 0.055 for fusion.

The absolute low-FPR recall is set by the most lens-like unrelated multiple-merger events, where the posterior branch supplies a source-consistency test complementary to waveform morphology.
This gain should be read as an averaged held-out lens-family effect rather than a uniform improvement at every fixed-FPR operating point.
As shown below, the \(1\%\)-FPR recall decreases slightly for PM and SIS while increasing for SIE and SIE+Shear, so the fixed-FPR behavior is controlled by the lens-family-dependent tail of the unrelated multiple-merger scores.
The deliberately low-dimensional posterior target removes timing, amplitude, and most residual morphology.
Waveform morphology carries most of the discrimination, while the posterior changes the ranking of ambiguous events.

We bootstrap the 28 held-out test conditions to quantify finite-test-condition uncertainty for the selected checkpoints.
The 95\% intervals are \(0.756\)--\(0.814\) for the direct ROC-AUC and \(0.831\)--\(0.857\) for the fusion ROC-AUC.
The paired fusion-minus-direct gain is \(0.058\) with interval \(0.032\)--\(0.086\) in ROC-AUC and \(0.053\) with interval \(0.021\)--\(0.084\) in PR-AUC.
For the \(0.1\%\) FPR recall, the paired gain is \(0.035\), with interval \(0.015\)--\(0.057\).
The \(1\%\) FPR recall gain is smaller and its interval crosses zero, consistent with the unrelated multiple-merger tail behavior discussed below.
Table~\ref{tab:overall} summarizes these averaged held-out metrics.

\begin{table*}[t]
\caption{
Average performance over all physical lens families and SNR values.
The numbers are ranking diagnostics for the fixed validation-selected checkpoints and simulation priors used in this study.
They are not calibrated catalog-level detection probabilities.
Condition-bootstrap intervals over the 28 lens-family/SNR conditions are reported in the text.
}
\label{tab:overall}
\begin{ruledtabular}
\begin{tabular}{lcccc}
Model & ROC-AUC & PR-AUC & R@1\% FPR & R@0.1\% FPR \\
\hline
Direct waveform & 0.785 & 0.753 & 0.136 & 0.020 \\
Posterior set & 0.592 & 0.593 & 0.048 & 0.016 \\
Fusion & 0.844 & 0.805 & 0.155 & 0.055 \\
\end{tabular}
\end{ruledtabular}
\end{table*}

\begin{figure*}[t]
\centering
\includegraphics[width=0.82\textwidth]{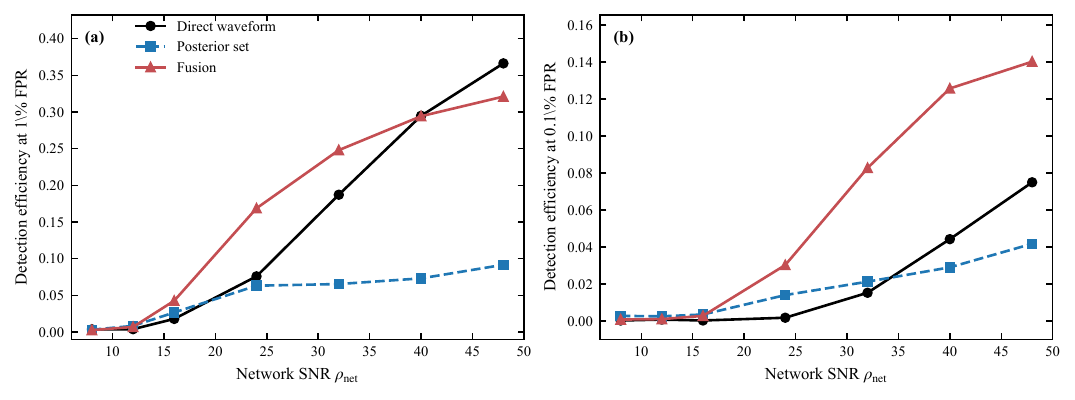}
\caption{
SNR-dependent detection efficiency averaged over the four physical lens families.
(a) Efficiency at the \(1\%\) reference-FPR threshold.
(b) Efficiency at the \(0.1\%\) reference-FPR threshold.
}
\label{fig:ranking_curves}
\end{figure*}

The SNR dependence is more transparent at fixed false-positive thresholds.
Figure~\ref{fig:ranking_curves} shows detection efficiency as a function of network SNR at \(1\%\) and \(0.1\%\) reference FPR.
The main improvement occurs at intermediate SNR.
For example, after averaging over the four held-out lens families, the \(1\%\)-FPR recall changes from 0.018 to 0.043 at \(\rho_{\rm net}=16\), from 0.076 to 0.169 at \(\rho_{\rm net}=24\), and from 0.187 to 0.248 at \(\rho_{\rm net}=32\).
At lower SNR, both scores remain below the rare-event threshold.
At high SNR, waveform morphology becomes sufficiently clear that the direct classifier partially catches up at the looser threshold.

\subsection{Physical information in source-consistency posteriors}

\begin{figure*}[t]
  \centering
\includegraphics[width=0.86\textwidth]{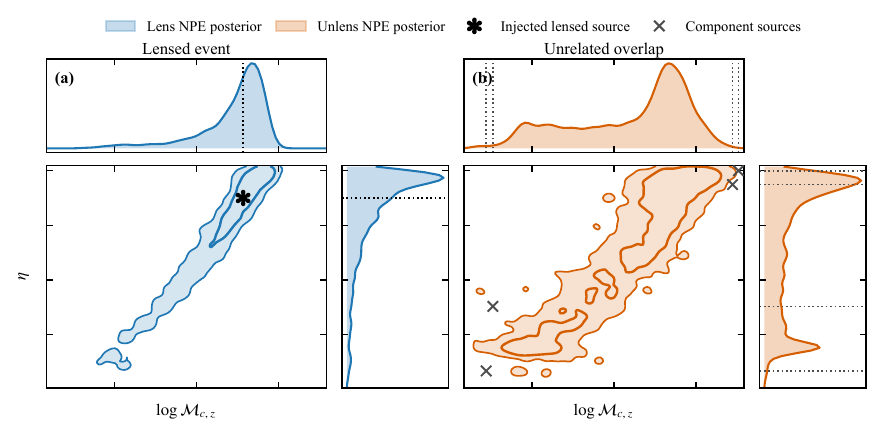}
  \caption{
Source-consistency posterior examples.
(a) Lensed validation event in the \((\log \mathcal{M}_{c,z},\eta)\) plane.
(b) Unrelated multiple-merger event evaluated under the same common-source posterior model.
The lower panels show smoothed 68.3\% and 95.5\% highest-posterior-density contours, and the upper and right panels show the corresponding one-dimensional marginal densities.
In panel (b), gray crosses and dotted guide lines mark the detector-frame \((\log \mathcal{M}_{c,z},\eta)\) values of the independent component mergers.
They are source markers, not class labels.
  }
\label{fig:posterior_contours}
\end{figure*}

The posterior branch provides an interpretable common-source diagnostic.
The pair \((\log \mathcal{M}_{c,z},\eta)\) preserves the leading mass-dependent phase-evolution information, while timing, amplitude, detector response, and residual morphology remain available to the waveform branch.
Figure~\ref{fig:posterior_contours} visualizes the contrast: a lensed event yields a compact posterior near the injected source parameters, whereas an unrelated multiple-merger event can broaden or split the forced common-source posterior.

\begin{figure*}[t]
\centering
\includegraphics[width=0.88\textwidth]{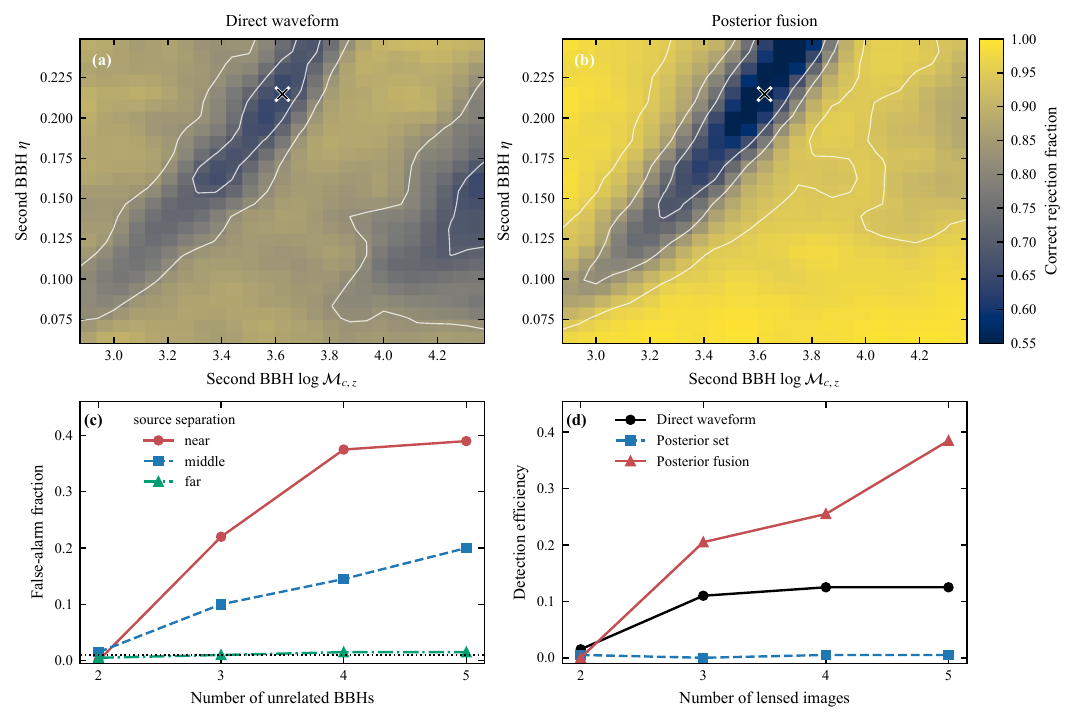}
\caption{
Controlled source-consistency tests.
(a,b) Correct rejection fraction for two unrelated BBHs at a threshold that gives \(50\%\) detection efficiency for SIE lensed events.
One BBH is fixed at the marked source, and the other is scanned across the \((\log \mathcal{M}_{c,z},\eta)\) plane.
The two panels use the same grid and the same color scale, with 100 examples per cell.
The contours are smoothed iso-rejection curves.
(c) Fusion false-alarm fraction as the number of unrelated BBHs increases, separated by source-parameter distance from the anchor.
(d) Detection efficiency for lensed signals as the number of lensed images changes in the matched controlled samples.
}
\label{fig:unlensed_factors}
\end{figure*}

The same source-consistency idea can be tested on a controlled grid.
In Fig.~\ref{fig:unlensed_factors}, one BBH is fixed at an anchor source and the second BBH is scanned across the \((\log \mathcal{M}_{c,z},\eta)\) plane.
Panels (a) and (b) use a threshold chosen to recover \(50\%\) of the SIE lensed events.
This threshold is used only to expose the morphology of difficult unrelated multiple-merger events.
The plotted quantity is the correct rejection fraction for unrelated two-BBH multiple-merger events.
Low values therefore mark regions where an unrelated multiple-merger event is classified as lensed.
The direct waveform map contains broad low-rejection regions away from the fixed source, showing that waveform morphology alone can mistake some unrelated multiple-merger events for lensed signals over a sizable part of the mass plane.
The fusion map rejects most of these broad regions and leaves a narrower low-rejection band that passes through the fixed source and follows the expected mass-ratio degeneracy direction.
This behavior is expected for a source-consistency test.
It helps away from the common-source degeneracy, but it cannot remove unrelated multiple-merger events that are genuinely close to a shared mass-plane posterior.
For the fusion classifier, the false-lensed fraction is 0.141 near the anchor and 0.014 in the far-source control.
Panels (c) and (d) show that the remaining failures are controlled by source similarity and multiplicity.
False alarms rise when more unrelated BBHs occupy the same window and when their source parameters are close to the fixed anchor.
For lensed events, the detection efficiency also increases with image number.
These two trends are not contradictory.
Panel (c) is an unrelated multiple-merger false-alarm test, where more unrelated mergers provide more chances to mimic a common source.
Panel (d) is a lensed-event efficiency test, where more lensed images provide more repeated-image evidence.

These controlled tests diagnose the difficult unrelated multiple-merger mechanism rather than astrophysical event rates.
For a detected BBH rate \(R\), the expected number of additional unrelated mergers in a \(\Delta t\simeq0.6\,{\rm s}\) window is only \(R\Delta t\).

\begin{figure*}[t]
\centering
\includegraphics[width=0.90\textwidth]{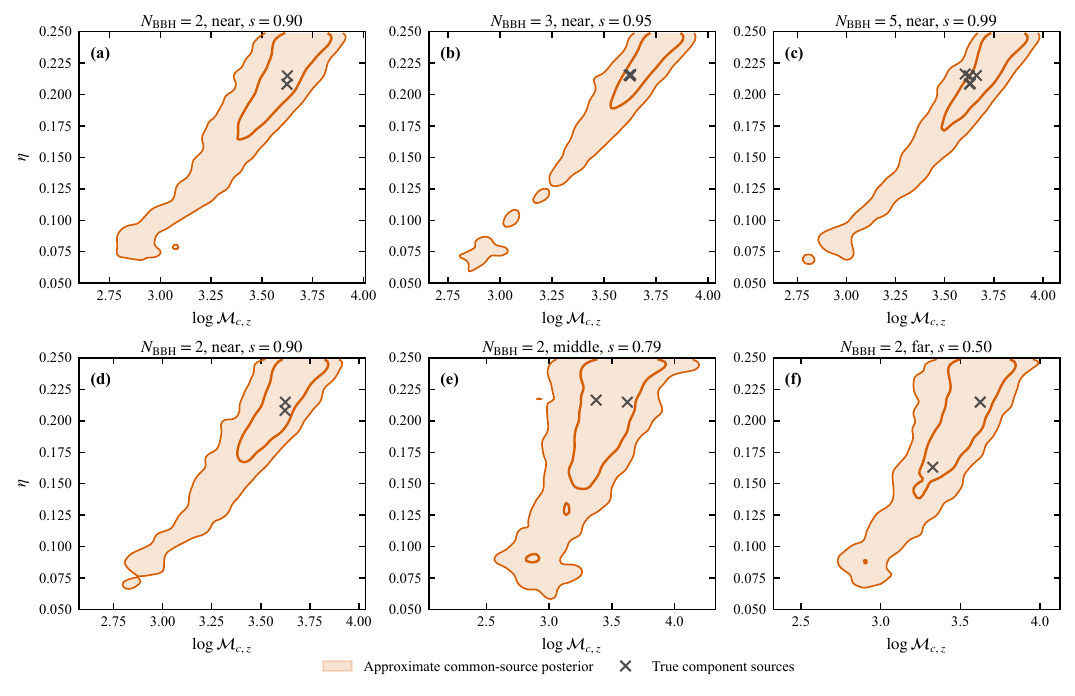}
\caption{
Controlled posterior contours for unrelated multiple-merger events.
The orange regions show the 68.3\% and 95.5\% highest-posterior-density contours of the simulation-trained approximate common-source posterior.
Gray crosses mark the true component-source locations in the \((\log \mathcal{M}_{c,z},\eta)\) plane.
The first row varies the number of nearby component BBHs.
The second row fixes \(N_{\rm BBH}=2\) and varies the source-parameter separation.
The title of each panel reports the fusion score \(s\), where larger values are more lens-like.
}
\label{fig:controlled_posteriors}
\end{figure*}

Figure~\ref{fig:controlled_posteriors} shows the same mechanism at the posterior level.
The first row keeps the unrelated sources close in the mass plane and increases the number of component mergers.
The posterior remains aligned with the mass-ratio degeneracy direction, and the fusion score rises as more source markers crowd the same region.
The second row fixes the number of component mergers to two and changes only their source-parameter separation.
The near case is the most difficult unrelated multiple-merger event, while the far case receives a much lower fusion score.
These examples explain why the false-alarm tail is not made of arbitrary event superpositions.
It is dominated by unrelated multiple-merger events that are close enough to one common-source posterior to survive the physical consistency check.

\begin{figure*}[t]
\centering
\includegraphics[width=0.78\textwidth]{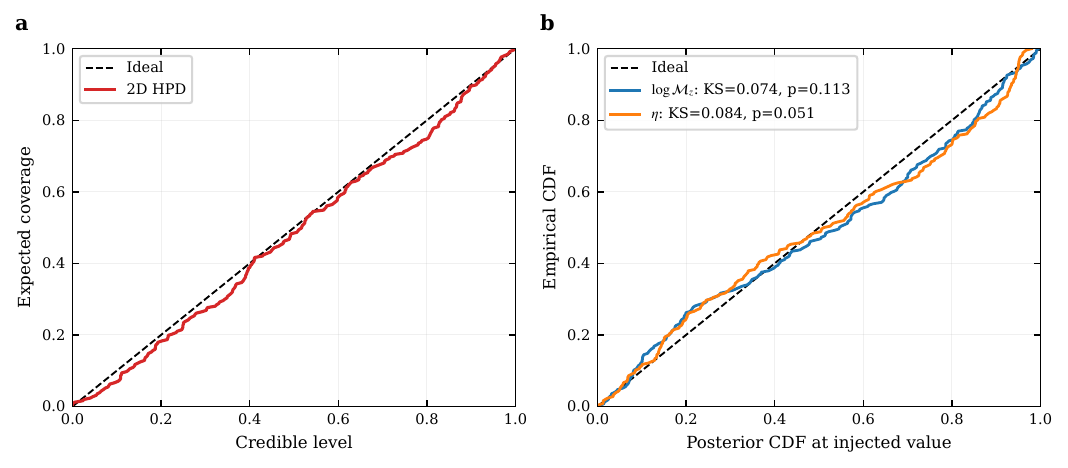}
\caption{
Reliability diagnostics for the source-posterior estimator on 256 validation events.
(a) Joint coverage test for the two-dimensional highest-posterior-density credible regions: the horizontal axis is the nominal credible level and the vertical axis is the empirical fraction of injections enclosed by the corresponding region.
(b) Marginal probability--probability diagnostics for \(\log \mathcal{M}_{c,z}\) and \(\eta\): the horizontal axis is the posterior cumulative distribution function (CDF) evaluated at the injected value and the vertical axis is the empirical CDF of these probability-integral-transform values.
The one-sample Kolmogorov--Smirnov tests in panel (b) check marginal uniformity, whereas panel (a) checks joint credible-region coverage.
The marginal KS statistics are \(D=0.074\) with \(p=0.113\) for \(\log \mathcal{M}_{c,z}\), and \(D=0.084\) with \(p=0.051\) for \(\eta\).
}
\label{fig:sbi_pp}
\end{figure*}

The reliability diagnostics in Fig.~\ref{fig:sbi_pp} show that the learned density behaves as an uncertainty estimate for this two-dimensional target under the simulation model.
Together, Figs.~\ref{fig:posterior_contours}--\ref{fig:sbi_pp} explain why the posterior-set branch is weak alone but useful in fusion.
It discards much of the waveform morphology and contributes source-consistency information.
These diagnostics support the physical interpretation of the fusion gain.
The main test of the method is the held-out evaluation on physical lens families.

\begin{figure*}[t]
\centering
\includegraphics[width=0.88\textwidth]{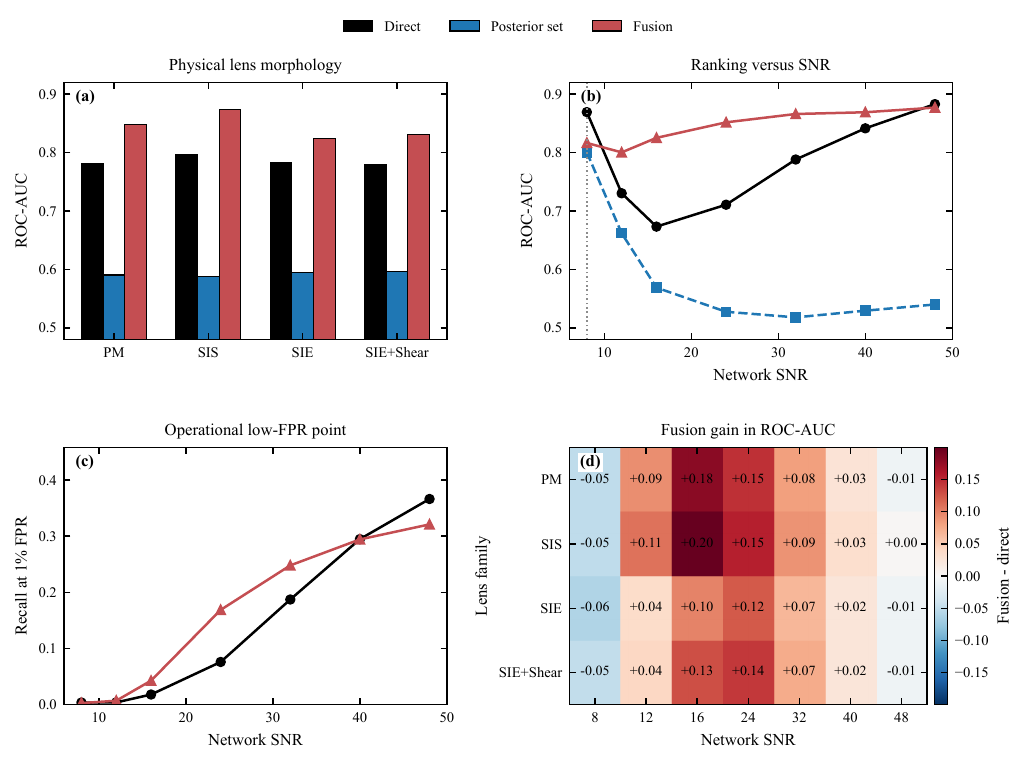}
\caption{
Curve-based held-out lens-family performance for the SNR-matched evaluation.
(a) ROC-AUC for different physical lens families.
(b) ROC-AUC as a function of network SNR.
(c) Recall at \(1\%\) FPR as a function of network SNR for the direct and fusion classifiers.
(d) ROC-AUC gain of fusion over the direct classifier across lens family and SNR.
}
\label{fig:morphology_variables}
\end{figure*}

Fusion gains are not confined to one lens family and are largest where the waveform-only boundary is least stable, especially at intermediate SNR.
Figure~\ref{fig:morphology_variables} separates the same held-out behavior by lens family, SNR, and fixed-FPR recall.

Table~\ref{tab:lens_family} shows that fusion gives the best ROC-AUC and PR-AUC for all four physical lens families, with the largest ROC-AUC gain for SIS, from 0.797 to 0.873.
The posterior-set model is nearly flat across families, with ROC-AUC values between 0.587 and 0.597, as expected for a lens-family-agnostic source-consistency diagnostic.
Family dependence mainly appears at fixed-FPR thresholds, where the score is controlled by the extreme unrelated multiple-merger tail.
At \(1\%\) FPR, fusion is slightly below the direct classifier for PM and SIS but is higher for SIE and SIE+Shear.
At \(0.1\%\) FPR the largest gains again come from SIE-family tests.
The data therefore do not support a blanket statement that fusion improves every rare-event operating point.
They show instead that source-consistency information is most useful in the SIE-family tests.
The stronger low-FPR gains for SIE and SIE+Shear are consistent with ellipticity and external shear broadening the image configurations, making source consistency less tied to a particular image geometry.

\begin{table*}[t]
\caption{
Performance averaged over the sampled SNR range for each physical lens family.
These are fixed-checkpoint point estimates.
The overall condition-bootstrap uncertainty is reported with Table~\ref{tab:overall}.
}
\label{tab:lens_family}
\begin{ruledtabular}
\begin{tabular}{llcccc}
Lens family & Model & ROC-AUC & PR-AUC & R@1\% FPR & R@0.1\% FPR \\
\hline
PM & Direct waveform & 0.781 & 0.746 & 0.124 & 0.015 \\
PM & Posterior set & 0.591 & 0.589 & 0.047 & 0.017 \\
PM & Fusion & 0.848 & 0.806 & 0.117 & 0.017 \\
SIS & Direct waveform & 0.797 & 0.761 & 0.147 & 0.014 \\
SIS & Posterior set & 0.587 & 0.589 & 0.045 & 0.015 \\
SIS & Fusion & 0.873 & 0.827 & 0.127 & 0.019 \\
SIE & Direct waveform & 0.784 & 0.753 & 0.134 & 0.023 \\
SIE & Posterior set & 0.595 & 0.596 & 0.049 & 0.016 \\
SIE & Fusion & 0.824 & 0.791 & 0.179 & 0.084 \\
SIE+Shear & Direct waveform & 0.780 & 0.751 & 0.138 & 0.026 \\
SIE+Shear & Posterior set & 0.597 & 0.598 & 0.048 & 0.017 \\
SIE+Shear & Fusion & 0.830 & 0.798 & 0.197 & 0.100 \\
\end{tabular}
\end{ruledtabular}
\end{table*}

\subsection{Operating-point diagnostics and low-false-alarm behavior}

\begin{figure}[!tbp]
\centering
\includegraphics[width=\columnwidth]{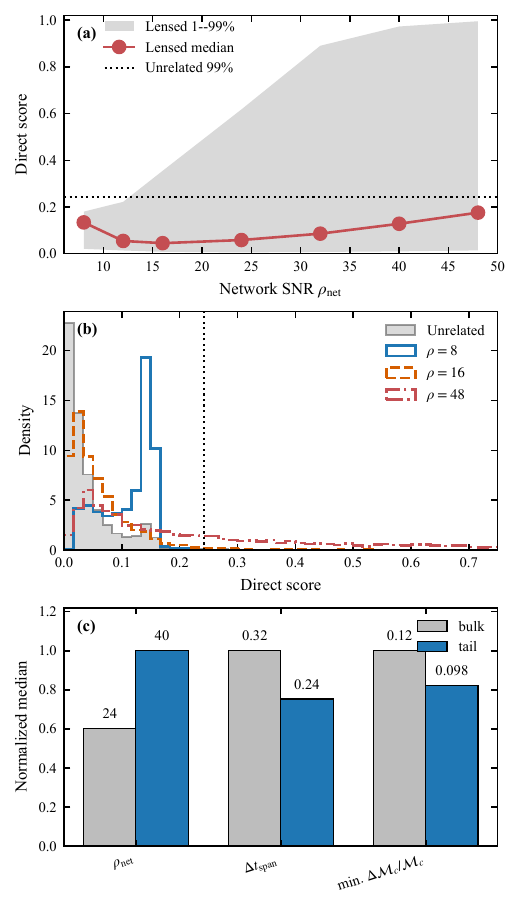}
\caption{
Direct-classifier diagnostic for the nonmonotonic SNR trend.
(a) Lensed-score quantiles as a function of SNR compared with the unrelated multiple-merger \(99\%\) score threshold that sets the \(1\%\)-FPR operating point.
(b) Per-example score distributions for selected SNR values.
(c) The top \(1\%\) unrelated multiple-merger score tail is louder, more temporally compact, and slightly closer in nearest-neighbor chirp mass than the bulk unrelated multiple-merger sample.
The bars are normalized by the larger median in each pair, with raw medians printed above the bars.
}
\label{fig:direct_snr_diagnostic}
\end{figure}

The nonmonotonic direct-classifier trend in Figs.~\ref{fig:morphology_variables} and \ref{fig:direct_snr_diagnostic} reflects the difference between global ranking and rare-event ranking.
Because both hypotheses contain multiple compact-binary waveforms, increasing SNR sharpens structure in both classes, and fixed-FPR recall is controlled by the high-score tail of unrelated multiple-merger events rather than by the average class separation measured by ROC-AUC.

The \(\rho_{\rm net}=8\) direct point is stable under five independent unrelated multiple-merger batches, giving \({\rm ROC\mbox{-}AUC}=0.866\pm0.008\) but only \(0.005\pm0.002\) recall at \(1\%\) FPR.
The score distribution explains this separation: the lensed-event median and \(99\%\) quantile are about 0.134 and 0.180, below the unrelated multiple-merger tail threshold of about 0.247 that sets the \(1\%\)-FPR operating point.
Thus low-SNR lensed events outrank the bulk of the unrelated multiple-merger events but rarely exceed the most lens-like tail events.

The difficult unrelated multiple-merger tail has a clear physical character: compared with the remaining unrelated multiple-merger events, the top \(1\%\) of unrelated multiple-merger scores have higher median SNR (40 versus 24), shorter median coalescence-time span (\(0.244\,{\rm s}\) versus \(0.324\,{\rm s}\)), and smaller nearest-neighbor relative chirp-mass separation (0.098 versus 0.120).
The rare-event tail is therefore the loud, temporally compact, and partly source-consistent subset of unrelated multiple-merger events.

The population-shift test makes this operating-point issue explicit.
When the evaluation population is changed to an observationally motivated BBH mass population, the balanced ROC-AUC values are 0.591 for the direct classifier and 0.605 for the main mass-plane fusion model, but thresholds calibrated on the original reference unrelated multiple-merger sample no longer represent their nominal false-alarm rates.
The empirical false-alarm probabilities become 0.125 and 0.260 for direct and mass-plane fusion at the nominal \(1\%\) reference-FPR threshold, and 0.018 and 0.140 at the nominal \(0.1\%\) threshold.
Thus the low-FPR numbers in this paper should be interpreted as diagnostics under specified simulated unrelated multiple-merger populations, not as catalog-level false-alarm probabilities.
A practical search must construct the unrelated multiple-merger population for the target catalog and report detection efficiency at that calibrated threshold.

\begin{figure}[!tbp]
\centering
\includegraphics[width=\columnwidth]{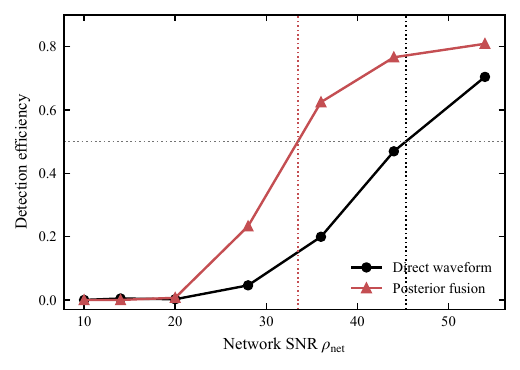}
\caption{
Population efficiency as a function of network SNR for the observationally motivated BBH mass population at the nominal \(1\%\) reference threshold.
Vertical dotted lines mark the interpolated SNR values at \(50\%\) detection efficiency.
}
\label{fig:snr_equivalent_d50}
\end{figure}

Figure~\ref{fig:snr_equivalent_d50} translates the same population test into a more physical efficiency scale.
The injections have luminosity distances, but this test rescales each waveform to a chosen SNR range.
The distance bins are therefore decoupled from detectability and should not be read as a true Mpc sensitive distance.
A true distance reach could be obtained from a non-rescaled injection campaign, or from the stored pre-rescaling SNR and scale factors together with a specified detector sensitivity and selection function.
We therefore quote only an SNR-equivalent \(D_{50}\) scale for the present evaluation.
At the nominal \(1\%\) reference threshold, the direct classifier reaches \(50\%\) detection efficiency at \(\rho_{\rm net}=45.3\), while the fusion classifier reaches the same efficiency at \(\rho_{\rm net}=33.5\).
Since distance scales approximately as \(D\propto1/\rho\) at fixed source orientation and mass, this corresponds to \(D_{50}^{\rm SNR}\) larger by a factor of 1.35 for the fusion ranking.
This is not a rate forecast, but it gives a physical reading of the ranking gain.

Posterior information contributes most where waveform-only ranking is fragile.
High-SNR morphology is already cleaner, while the lowest-SNR stringent-FPR performance remains controlled by the unrelated multiple-merger tail.
The largest fusion gains therefore appear at intermediate SNR and in several SIS/SIE-family tests.

The classifier sees only the generic multi-image construction during training and is tested afterward on PM, SIS, SIE, and SIE+Shear lenses without labels or parameter targets tied to these lens potentials.
Within this geometric-optics macrolens scope, the two-parameter posterior improves average event ranking without lens-family labels.
Substructure, wave-optics effects, detector nonstationarity, and broader unrelated multiple-merger populations can change the decision boundary.

\medskip
\noindent\textit{Detector-response consistency.}

We also performed a controlled detector-response extension in which the posterior target is enlarged to \((\log \mathcal{M}_{c,z},\eta,\alpha,\sin\delta)\).
This test should not be interpreted as adding new intrinsic phase-evolution information.
Sky position affects the detector response through inter-site time delays and antenna-pattern projection, not the chirp-frequency evolution of the binary.
The extension keeps the average fusion ROC-AUC close to the main result, \(0.839\) instead of \(0.844\), but increases the most stringent \(0.1\%\)-FPR recall from 0.055 to 0.089.
The added information enters through a stricter common-source requirement.
A difficult unrelated multiple-merger event must then be compatible not only in the mass plane but also in the network projection.
This is still a conservative H1/L1 test, since two-detector O4 data provide weak sky localization for sub-second images.

\begin{figure}[!tbp]
\centering
\includegraphics[width=\columnwidth]{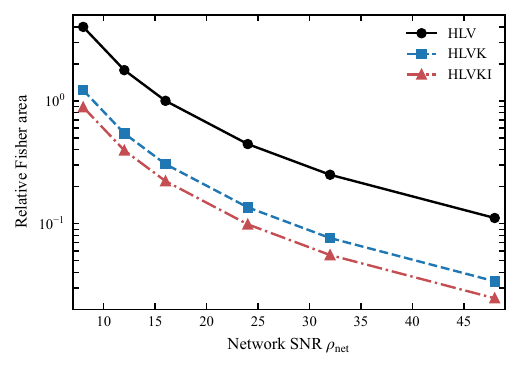}
\caption{
Timing-only Fisher estimate for detector-network localization.
The vertical axis is the relative sky area normalized to the HLV network at \(\rho_{\rm net}=16\).
The estimate assumes an effective bandwidth of \(100\,{\rm Hz}\) and uses timing precision only.
}
\label{fig:detector_network_localization}
\end{figure}

To estimate how this information could scale with detector number, we used a timing-only Fisher estimate for Hanford--Livingston--Virgo (HLV), Hanford--Livingston--Virgo--KAGRA (HLVK), and Hanford--Livingston--Virgo--KAGRA--LIGO--India (HLVKI) networks \cite{Wen:2010cr,Fairhurst:2010is}.
The timing uncertainty is taken as \(\sigma_t=(2\pi\rho_{\rm net}\sigma_f)^{-1}\), with an effective bandwidth \(\sigma_f=100\,{\rm Hz}\), and the reported area is normalized to HLV at \(\rho_{\rm net}=16\).
It is not a full sky-localization calculation.
It does not include antenna-pattern degeneracies, calibration uncertainties, inclination--distance degeneracy, or sky priors.
We do not assign a finite two-dimensional Fisher area to the two-detector HL network in this timing-only comparison, because one independent arrival-time delay leaves an annular degeneracy on the sky.
Within this restricted estimate, adding KAGRA and LIGO--India reduces the relative localization area at fixed SNR.
At \(\rho_{\rm net}=16\), the relative area is 1 by construction for HLV, 0.305 for HLVK, and 0.222 for HLVKI.
At \(\rho_{\rm net}=24\), the corresponding values are 0.444, 0.136, and 0.099.
Thus a larger network would make detector-response consistency more informative, within the limits of this Fisher scaling estimate.
For dark BBH lenses, this does not imply an electromagnetic transient counterpart by itself.
Improved localization could instead help identify a lens or host-galaxy environment when combined with galaxy catalogs and lensing information \cite{Chen:2025xeg}.
Figure~\ref{fig:detector_network_localization} shows the resulting relative localization-area scaling.

\section{Conclusion}
\label{sec:conclusion}

We have developed physics-informed posterior learning for ranking lensed multi-image signals against unrelated multiple-merger events.
The method encodes a simple geometric-optics fact in a learnable form.
Lensing changes amplitudes, arrival times, and Morse phase offsets, while preserving the intrinsic phase evolution of the source.
We represent this consistency through a simulation-trained approximate posterior over \((\log \mathcal{M}_{c,z},\eta)\), and fuse posterior samples with direct waveform features.

The tests support three conclusions.
First, incorporating the posterior representation into the fusion classifier improves event ranking on held-out PM, SIS, SIE, and SIE+Shear lenses, even though these lens families are not used as training labels.
Second, the gain is not a generic improvement at every operating point.
The difficult unrelated multiple-merger tail is made of loud, compact, and partly source-consistent unrelated multiple-merger events.
This explains why low-FPR performance is controlled by the tail of the unrelated multiple-merger distribution.
Third, the population test gives a physical scale for the improvement.
At the nominal \(1\%\) reference threshold, the detection efficiency rises from \(20.8\%\) to \(35.2\%\), and the SNR-equivalent \(D_{50}\) scale increases by a factor of 1.35.

This study remains a controlled nonspinning BBH simulation with H1/L1 O4 backgrounds.
Its thresholds should be calibrated with the unrelated multiple-merger population used in a real search.
The sky-extension test indicates where the next physical information should enter.
Two unrelated binaries can accidentally match in the mass plane, but a lensed interpretation must also be compatible with one detector response.
Future detector networks and repeated lensed images observed at different sidereal times can sharpen this consistency test, making gravitational-wave lensing a cleaner route from event ranking to lensing cosmology.

\section*{Acknowledgements}

This work was supported by the National Natural Science Foundation of China (Grants Nos. 12473001, 12503001, 12575049, and 12533001), the National SKA Program of China (Grants Nos. 2022SKA0110200 and 2022SKA0110203), the China Manned Space Program (Grant No. CMS-CSST-2025-A02), and the 111 Project (Grant No. B16009).
The data used in our study is based upon work supported by NSF's LIGO Laboratory, which is a major facility fully funded by the National Science Foundation.

\section*{Data Availability}

The processed plotting data, evaluation tables, metadata for the O4 background intervals, and documentation of the simulated data products supporting this study will be made publicly available upon acceptance through a public repository, with a persistent archive link supplied in the final version.

\clearpage

\appendix

\section{Training and evaluation details}
\label{app:training}

The SNR-matched evaluation used the best checkpoints of the source-posterior estimator, posterior-set classifier, fusion classifier, and direct waveform classifier.
The held-out lens-family evaluation used the physical-lens test directory and produced the aggregated table of generalization metrics used in this paper.
The SNR-matched lensed training pool consists of 100 independently generated files with \(10^4\) examples per file.
The unrelated multiple-merger pool uses the same file count and examples per file.
Thus the cyclic loader draws from \(10^6\) lensed examples and \(10^6\) unrelated multiple-merger examples before the within-file train/validation split is applied.
The source-posterior estimator uses only the lensed files because the target common-source parameters are defined only for the lensed class, whereas the posterior-set and fusion classifiers load one lensed file and one unrelated multiple-merger file in each training cycle.

The posterior estimator was trained for 100 epochs with batch size 4 in the low-memory run.
The posterior-set classifier was trained for 100 epochs with batch size 8 and 4 posterior samples per example.
The fusion classifier was trained for 100 epochs with batch size 4 and 4 posterior samples per example.
The evaluation used 32 posterior samples, direct batch size 64, posterior-set batch size 32, and fusion batch size 8.
All three evaluation branches used the same lensed and unrelated multiple-merger examples in each test condition.

\bibliography{paper}

\end{document}